\documentclass[review]{amsart}
\usepackage[utf8]{inputenc}
\usepackage{graphicx}
\usepackage{lineno}

\usepackage[english]{babel}
\usepackage{amssymb}
\usepackage{enumerate}
\usepackage{mathtools}
\usepackage{nicefrac}
\usepackage{graphicx}
\usepackage{amsmath}
\usepackage{subfigure}
\usepackage{booktabs}
\usepackage{tabularx}
\usepackage{verbatim}
\usepackage{adjustbox}
\usepackage{multirow}
\usepackage[margin=2cm]{geometry}
\RequirePackage[colorlinks,citecolor=blue,urlcolor=blue]{hyperref}
\usepackage{algorithm} \usepackage{algorithmic}
\usepackage{mathptmx}
\usepackage{latexsym}

\begin{document}

\begin{center}
{\bf{Desaturating EUV observations of solar flaring storms}}
\end{center}
\begin{center}
{\em{Sabrina Guastavino, Dipartimento di Matematica, Universit\`a di Genova, Genova, Italy}}
\end{center}
\begin{center}
{\em{Michele Piana, Dipartimento di Matematica, Universit\`a di Genova, Genova and CNR - SPIN, Genova, Italy}}
\end{center}
\begin{center}
{\em{Anna Maria Massone, Dipartimento di Matematica, Universit\`a di Genova, Genova and CNR - SPIN, Genova, Italy}}
\end{center}
\begin{center}
{\em{Richard Schwartz, NASA Goddard Space Flight Center, USA}}
\end{center}
\begin{center}
{\em{Federico Benvenuto, Dipartimento di Matematica, Universit\`a di Genova, Genova, Italy}}
\end{center}

%
%

%
%
%
%


{\bf{Image saturation has been an issue for several instruments in solar astronomy, mainly at EUV wavelengths. However, with the launch of the {\em{Atmospheric Imaging Assembly (AIA)}} as part of the payload of the {\em{Solar Dynamic Observatory (SDO)}} image saturation has become a big data issue, involving around $10^5$ frames of the impressive dataset this beautiful telescope has been providing every year since February 2010. This paper introduces a novel desaturation method, which is able to recover the signal in the saturated region of any {\em{AIA}} image by exploiting no other information but the one contained in the image itself. This peculiar methodological property, jointly with the unprecedented statistical reliability of the desaturated images, could make this algorithm the perfect tool for the realization of a reconstruction pipeline for {\em{AIA}} data, able to work properly even in the case of long-lasting, very energetic flaring events.}}

Extreme Ultraviolet (EUV) imaging is crucial for providing a clear-cut picture of the dynamical structure of the solar corona at many different time and spatial scales \cite{asetal01,beetal01,vaetal09,heetal11}. Observations at these wavelengths are probably the only data that can provide direct clear visualizations of magnetic reconnection as the trigger of magnetic energy release \cite{flhu01,yoetal01,qietal04,zhetal12}, reveal in detail the thermal structure of the solar atmosphere \cite{wabr09,hako12} and therefore explain basic plasma physics processes like coronal heating \cite{asetal08,asetal07} and irradiance \cite{woetal11}, and unveil still unresolved diagnostic issues concerned with coronal waves and oscillations \cite{naof01,lietal10}. From a space weather perspective, the ability of EUV imaging to point out, both spatially and dynamically, the connection between solar flares and coronal mass ejections (CMEs) paves the way to understand how Sun's variability impact the escape of energetic particles into the heliosphere \cite{maetal16,goetal18}. 

The {\em{Atmospheric Imaging Assembly}} \cite{leetal11} on board of the {\em{Solar Dynamics Observatory}} \cite{pe15} provides an unprecedented EUV view of the solar corona and its dynamics. With their multiple channels centered on seven EUV wavelengths ($94$ \AA, $131$ \AA, $171$ \AA, $193$ \AA, $211$ \AA, $304$ \AA, $335$ \AA) the four 
{\em{SDO/AIA}}  telescopes provide multiple simultaneous full-disk images of the corona and transition region above the solar limb, characterized by $1.5$ arcsec spatial resolution and $12$ second time cadence. These imaging performances allowed {\em{AIA}} to obtain several significant scientific results. Just as main examples, using this instrument it was possible to show that the total energy flux observed in low-frequency Alfven waves is sufficient to supply the energy heating of the quiet corona \cite{mcetal11}; to infer the temperature content in the prominence-corona transmission region \cite{paetal12}; to perform the first observation of transverse wave motions in solar polar plumes \cite{thetal14}. The use of {\em{SDO/AIA}} in flare observations, also in connection with data provided by the other two instruments in the payload of {\em{SDO}}, has provided impressive insights in at least three domains. First, this telescope has been able to picture the flare morphology and dynamics with an unprecedented level of details, up to the point to beautifully imaging the magnetic-reconnection in flaring loops of different classes \cite{taetal11,suetal13,tietal14}. Second, {\em{AIA}} images allowed differential emission measure analysis to quantitatively determine the temperature structure of the flaring loop \cite{bako12,peetal12}. Finally, the high spatio-temporal resolution of this instrument has allowed crucial insights in the phenomenological comprehension of pulsations and instabilities of the corona \cite{suetal12,doetal12,ofth11}.

\begin{figure}
\includegraphics[width=0.8\textwidth]{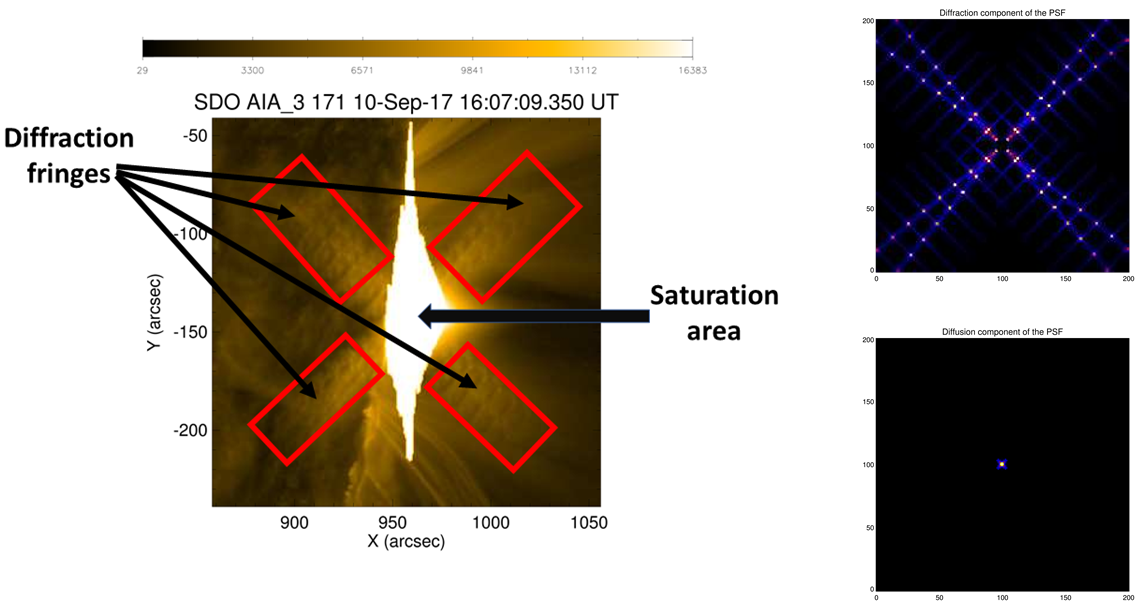}
\caption{Left panel: an example of saturated AIA image with highlighted the overall (primary + blooming) saturation region and the diffraction fringes; the event occurred on September, 10 2017 at the acquisition time 16:07:09 UT. Right-top panel: the diffraction component of the 171 $\mathring{A}$ AIA/SDO point-spread function; right-bottom panel: the diffusion component of the PSF.}\label{fig:figure-1} 
\end{figure}

As typically happens in EUV imaging, {\em{SDO/AIA}} observations of solar flares may be significantly limited by the presence of two kinds of imaging artifacts, diffraction and saturation, clearly visible in Figure \ref{fig:figure-1}, left panel. Diffraction is a direct consequence of the hardware characteristics of the {\em{AIA}} design. All telescopes introduce blur into the image formation process and such a blur is encoded in their Point Spread Function (PSF), modeling the impulse response to a point source placed far from the optical system. {\em{AIA}} PSF is characterized by the two components illustrated in the right panels of Figure \ref{fig:figure-1}: a core peak that induces diffusion effects and a regular, peripheral diffraction pattern of varying intensity that replicates the central peak due to wave scattering against the filters support. The diffraction component of the PSF correspondingly induces diffraction artifacts that typically show up simultaneously with saturation effects in the image core. Saturation is related to CCD-based imaging systems and is classified according to two different effects: primary saturation refers to the fact that, for intense incoming flux, CCD pixels loose their ability to accommodate additional charge; blooming, or secondary saturation, names the fact that primary saturation causes charge to spill into their neighbors. The resulting overall artifact tends to flatten and threshold the brightest core of these maps in a strongly anisotropic manner (north-south direction), thus degrading, both qualitatively and quantitatively, their imaging properties. 

Diffraction and saturation play a completely different, in some sense competing role in the image processing effort for {\em{AIA}}. Indeed, just part of the incoming signal accumulates in the CCD pixels up to saturation, while the other part is coherently and linearly scattered to produce diffraction pixels unaffected by saturation. As shown in \cite{scetal14,toetal15}, this fact has a crucial implication for image restoration: all information lost due to primary saturation is actually present, as regular ghosts, in the diffraction fringes and therefore mathematics can in principle help to recover those information by means of an inverse diffraction procedure. This approach has been implemented in the Solar SoftWare (SSW) tree by means of an IDL tool named {\em{DESAT}} \cite{scetal15}, made of three steps: the segmentation step utilizes thresholding and correlation in order to separate the input, saturated image into four regions (diffraction fringes, primary saturation, blooming, the remaining part of the map); the reconstruction step utilizes Expectation/Maximization \cite{lu74} in order to restore the information in the primary saturation region; the synthesis step projects the estimated flux onto the image space and glues it together with the image background. 

The three steps of the {\em{DESAT}} pipeline strongly exploit the knowledge of an estimate of the image background and this is the actual drawback of this approach. Background estimation is in general a tricky issue in solar imaging and {\em{DESAT}} addresses it by exploiting a specific aspect of {\em{AIA}} hardware. In fact, this telescope is equipped with a feedback system that automatically reduces the exposure time in correspondence of intense emission. It follows that a typical {\em{AIA}} observation along a time range of some minutes, is characterized by some unsaturated frames that can be utilized for background estimation. Specifically, for each saturated image, {\em{DESAT}} interpolates the pixel content belonging to the two unsaturated maps recorded just before and just after it and the resulting signal is assigned to the background pixels.

But what if {\em{AIA}} is observing an extremely intense flaring storm so that the feedback system becomes ineffective and strong saturation effects occur for a whole time series of acquired images? These dramatic events are very interesting for the physics of solar flares, for many different reasons: their intensity and duration make visible all possible morphological aspects of the emission; further, they are followed by a whole zoo of eruptive phenomena, from gamma burst, through coronal mass ejections to the generation of solar energetic particles; finally (and just for all these reasons) their impact on space weather may be particularly significant and therefore their full and detailed observation may lead to important discoveries in many areas of heliophysics. However, saturation makes the EUV observation of these monster flares almost useless, thus impeding scientists to open a crucial wavelength window over these highly interesting events. Take for example the September 2017 super-storm \cite{seda18}: between September 6-10, $27$ M flares and four X flares were emitted by the Sun, which correspondingly emitted several powerful CMEs and bursts of high energy protons. For more than one hour observation, all {\em{AIA}} filters suffered saturation in the core region of their images. In particular, at $171$ $\AA$ all images presented significant primary saturation and blooming effects, which corresponds to a consecutive deterioration of up to $300$ EUV maps: even a rather sophisticated computational method such as {\em{DESAT}} is ineffective in this case, since the background estimation via interpolation of unsaturated emission is completely impossible. 

The present study introduces a novel computational method for the analysis of a {\em{SDO/AIA}} saturated images able to recover the signal in the saturated region in a fast fashion without using any other information but the one contained in the image itself. As for {\em{DESAT}} the input data are represented by the diffraction fringes and therefore this is again an inverse diffraction algorithm. However, differently of {\em{DESAT}}, this new approach realizes desaturation and background estimation at the same time, by exploiting the energy information contained just in the image itself. The computational approach we formulated to obtain this result is described in Figure \ref{fig:figure-2} and is again made of three steps. In the first step the unknowns are the signal in the primary saturation region and the background, which for the moment is assumed as a constant offset to determine. Such unknowns are reconstructed by solving the constrained minimization problem of seeking the minimum number of pixels whose content reproduces the diffraction fringes and the overall flux recorded in the saturation region. To this aim we used a LASSO-type method, named PRiL, which accounts for the Poisson nature of the input observations \cite{gube18}. This first step also provides a first estimate of the background assumed here as a constant offset. Then, the second step implements an alternate running of Expectation Maximization and PRiL, whereby the unknowns are the background and the flux in the primary saturation region, respectively, and the input data are, correspondingly, the signal in the primary saturation and the background. This step is controlled by C-statistic: when the fitting of the diffraction fringes is satisfactory, the algorithm goes to step 3, which is devoted to the projection of the desaturated signal onto the image space and where the blooming region is reconstructed by means of a smoothing procedure \cite{ga10,waetal12}. 

\section*{Results}
\begin{figure}
\subfigure[{T16:05:45}]{
\includegraphics[width=0.2\textwidth, 
angle=90]{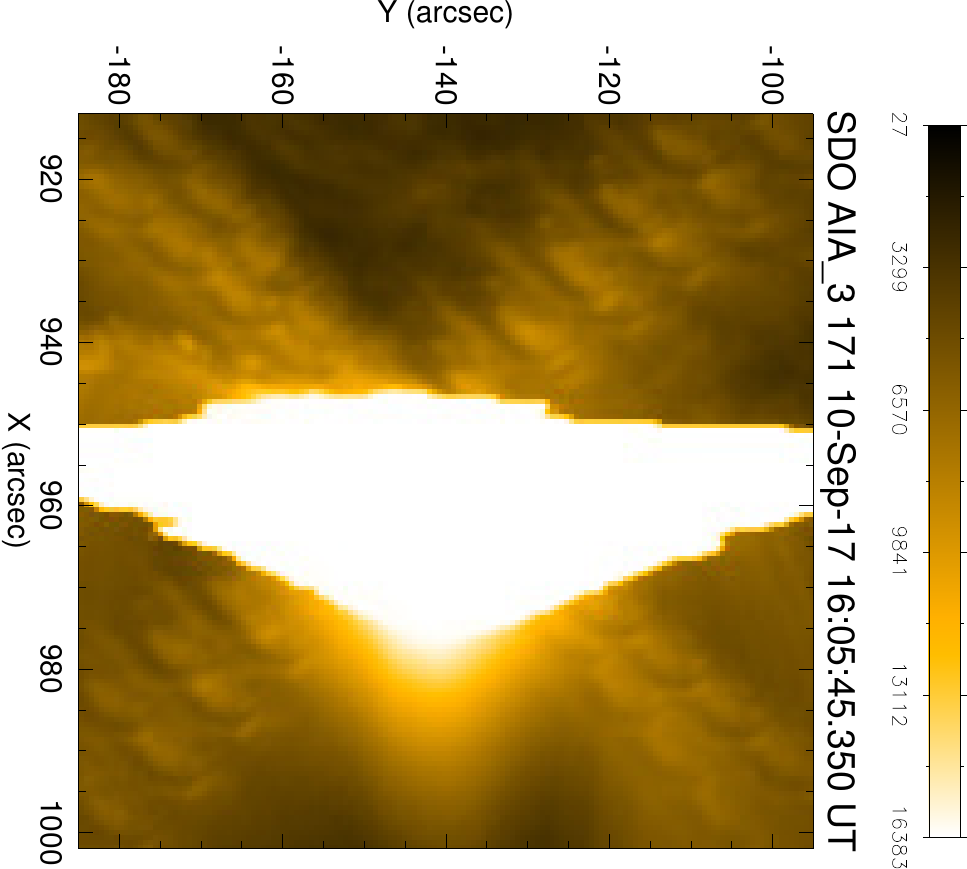}} 
\subfigure[{T16:05:57}]{\includegraphics[width=0.2\textwidth, angle=90]{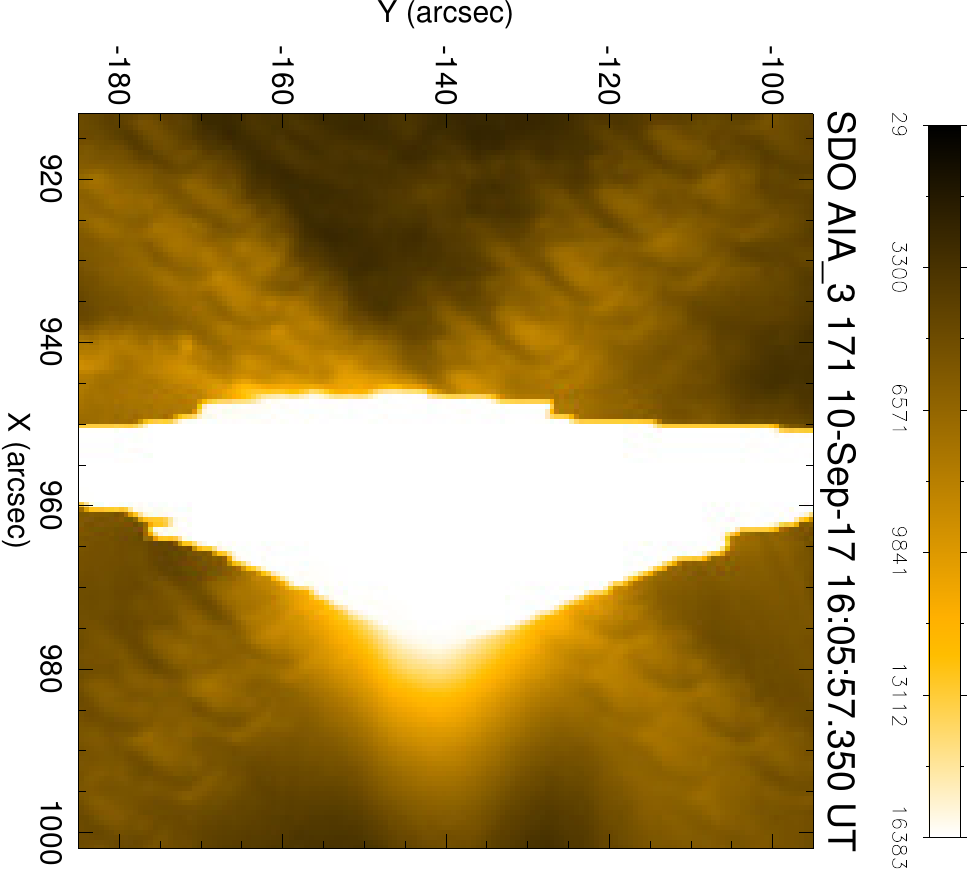}} 
\subfigure[{T16:06:09}]{\includegraphics[width=0.2\textwidth, angle=90]{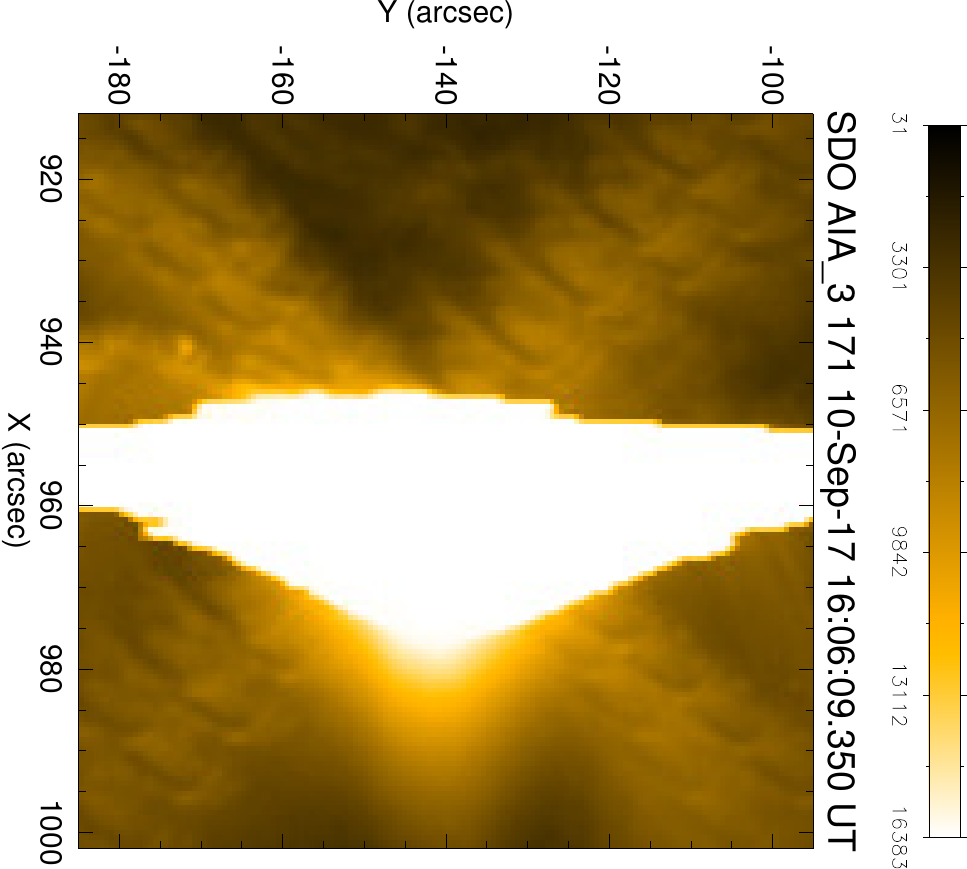}} 
\subfigure[{T16:06:21}]{\includegraphics[width=0.2\textwidth, angle=90]{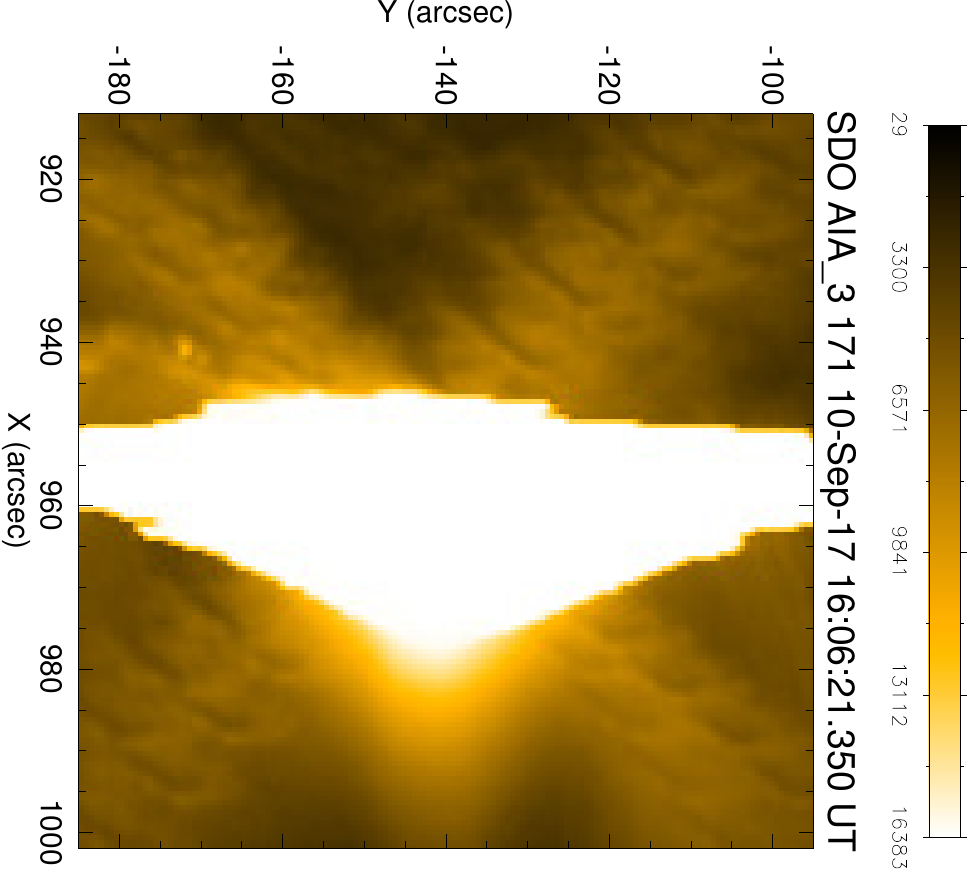}} 
\subfigure[{T16:06:33}]{\includegraphics[width=0.2\textwidth, angle=90]{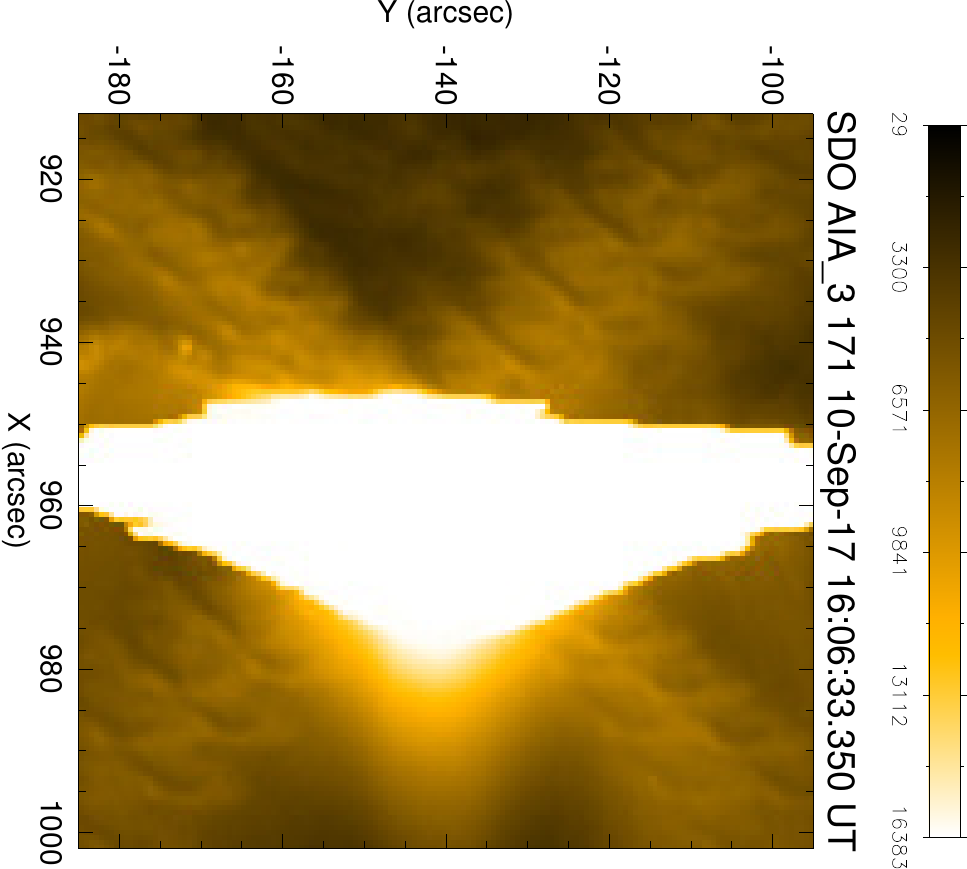}} 
\\
\subfigure[{T16:05:45}]{
\includegraphics[width=0.2\textwidth, angle=90]{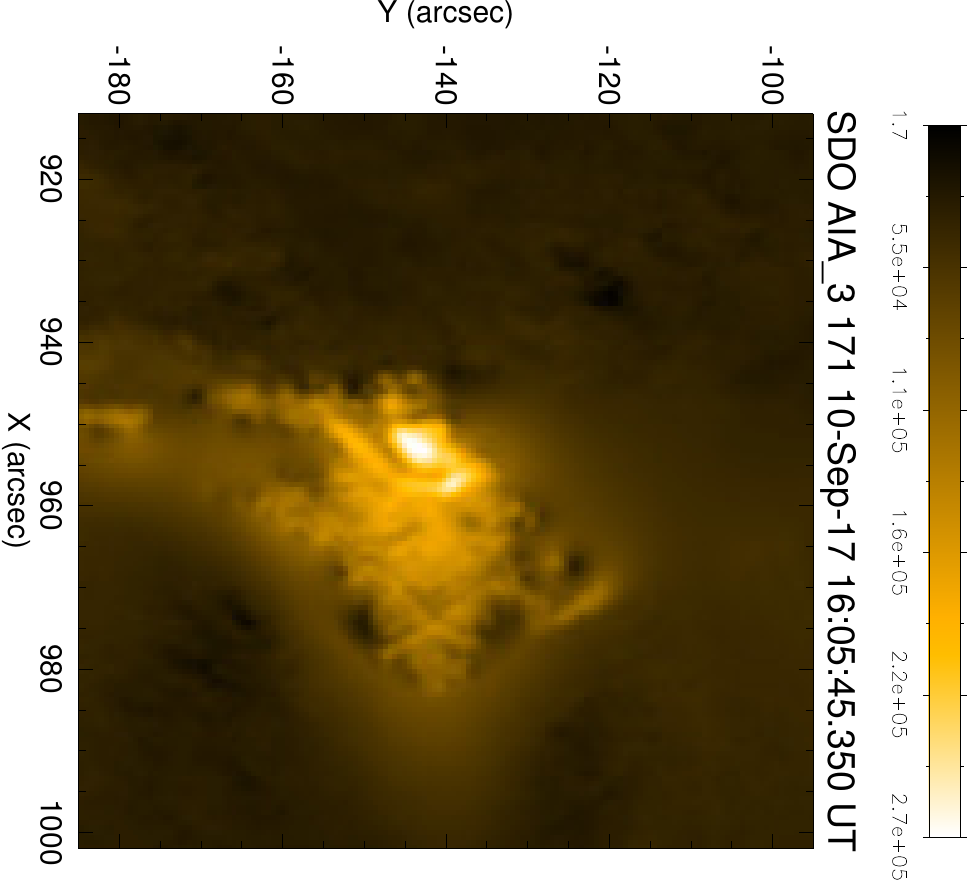}} 
\subfigure[{T16:05:57}]{\includegraphics[width=0.2\textwidth, angle=90]{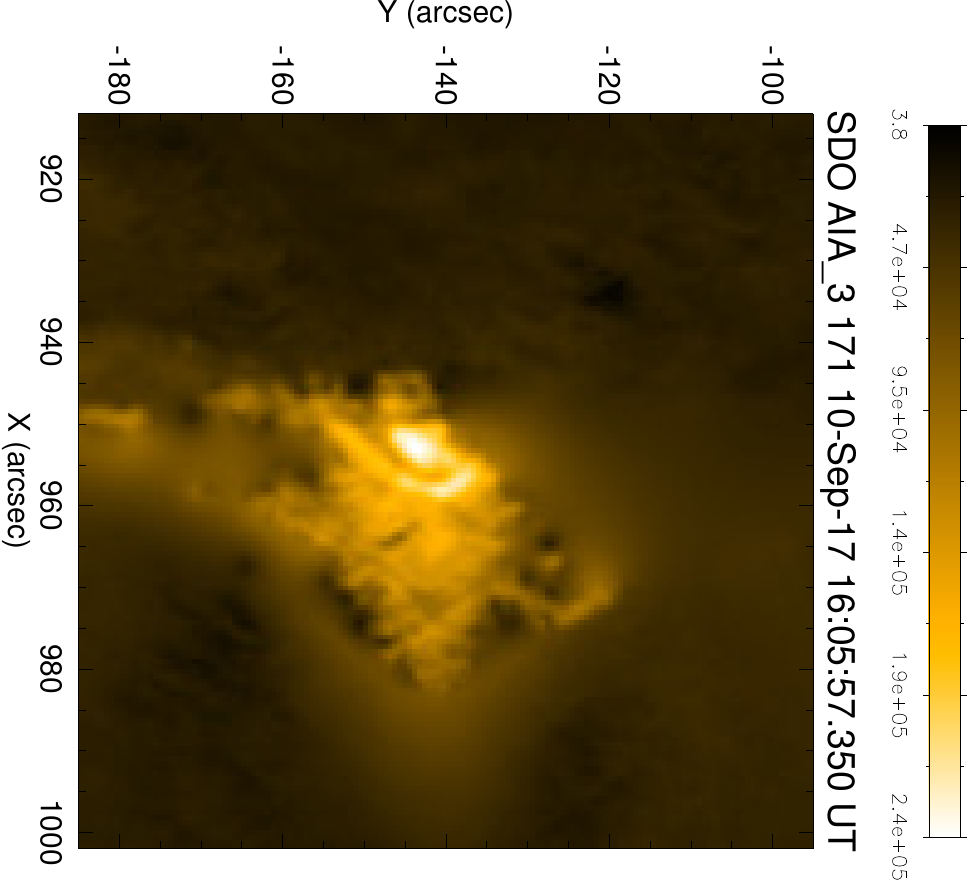}} 
\subfigure[{T16:06:09}]{\includegraphics[width=0.2\textwidth, angle=90]{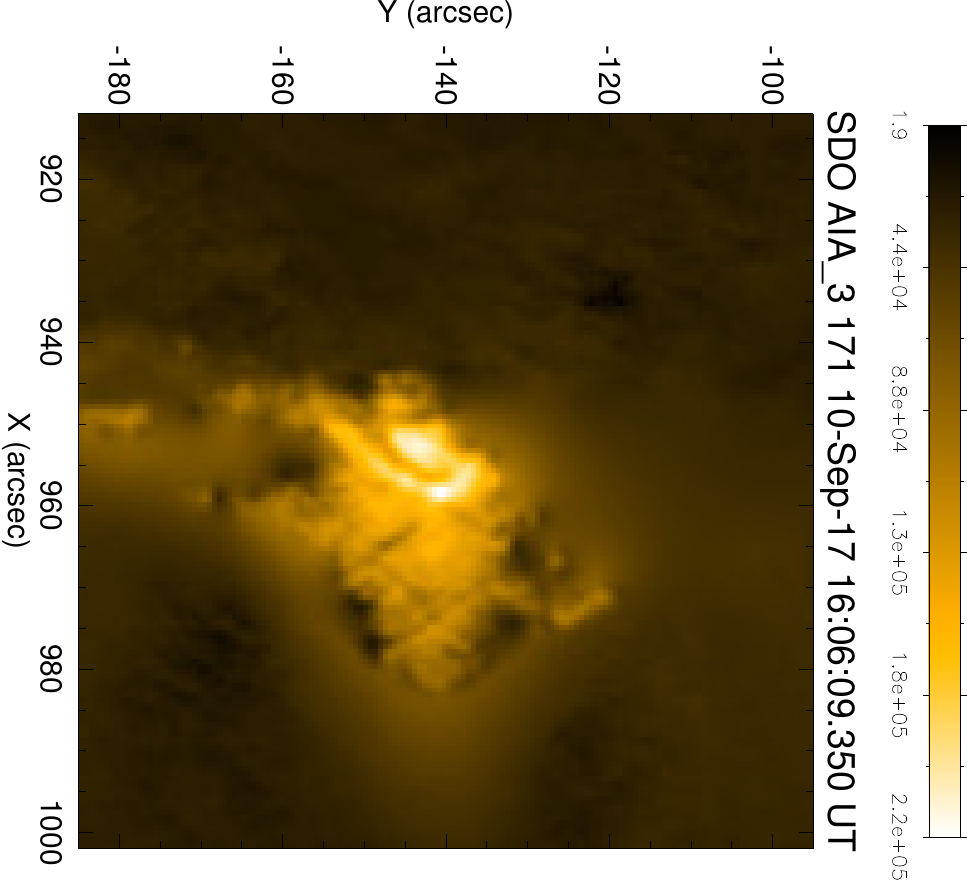}} 
\subfigure[{T16:06:21}]{\includegraphics[width=0.2\textwidth, angle=90]{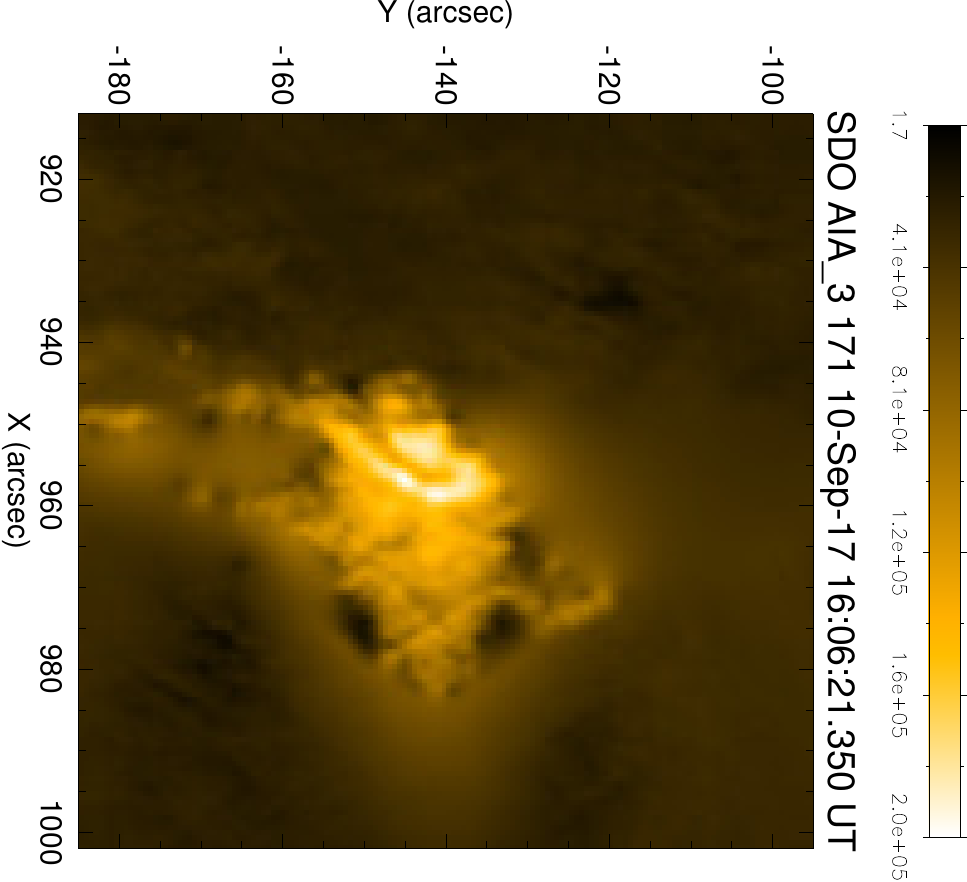}} 
\subfigure[{T16:06:33}]{\includegraphics[width=0.2\textwidth, angle=90]{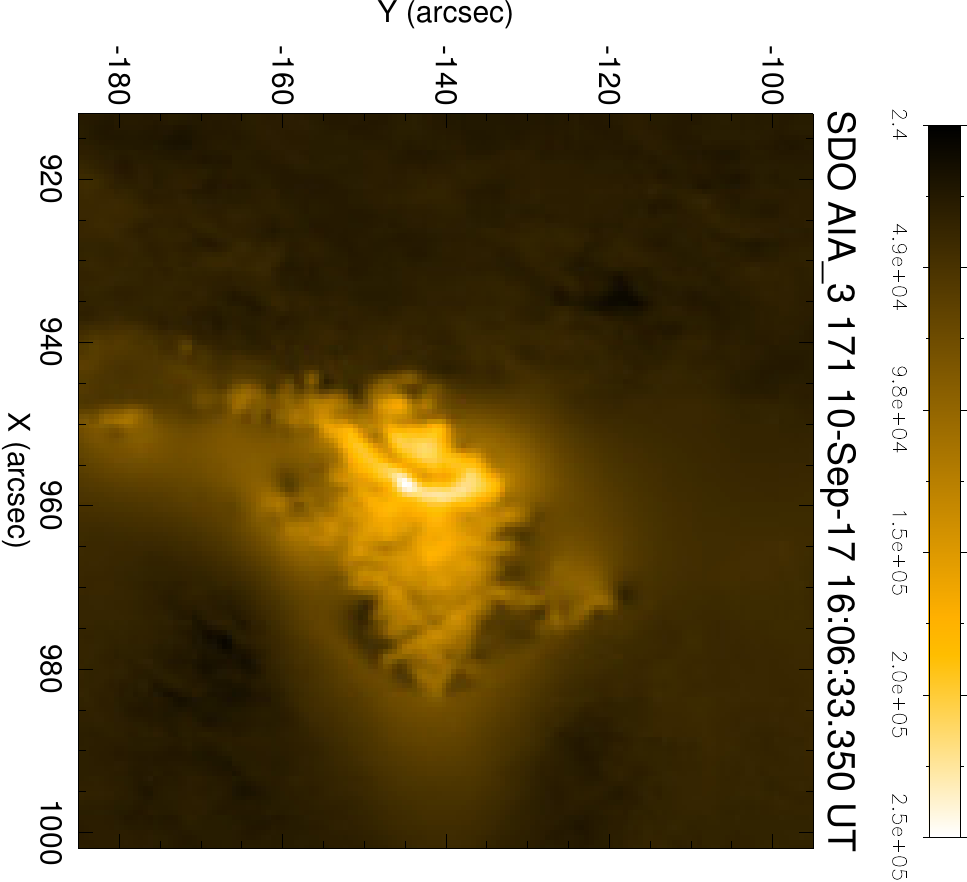}} 
\\

\subfigure[{T16:05:45}]{
\includegraphics[width=0.2\textwidth, angle=90]{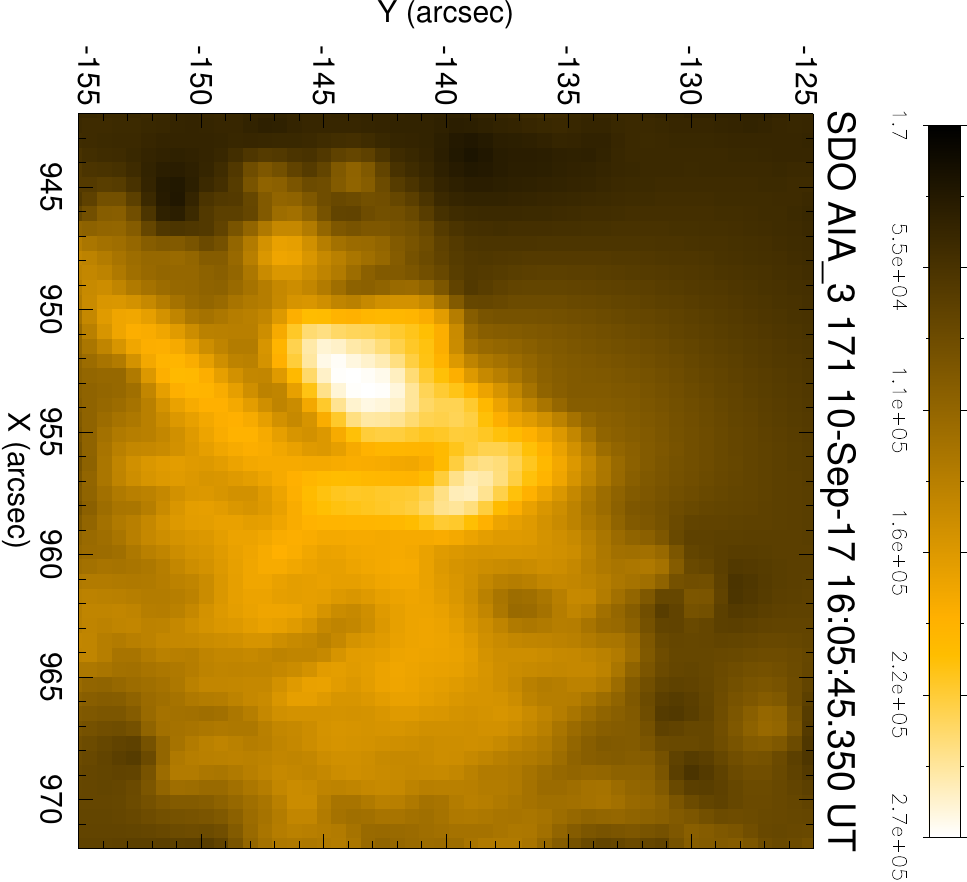}} 
\subfigure[{T16:05:57}]{\includegraphics[width=0.2\textwidth, angle=90]{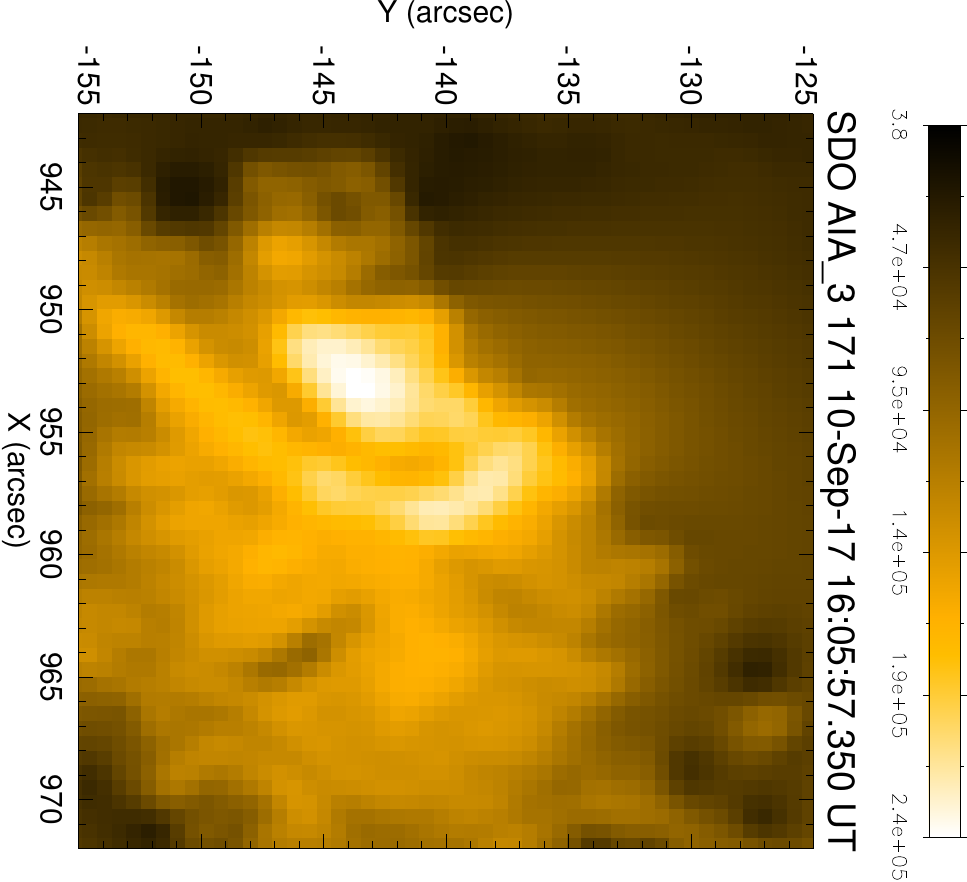}} 
\subfigure[{T16:06:09}]{\includegraphics[width=0.2\textwidth, angle=90]{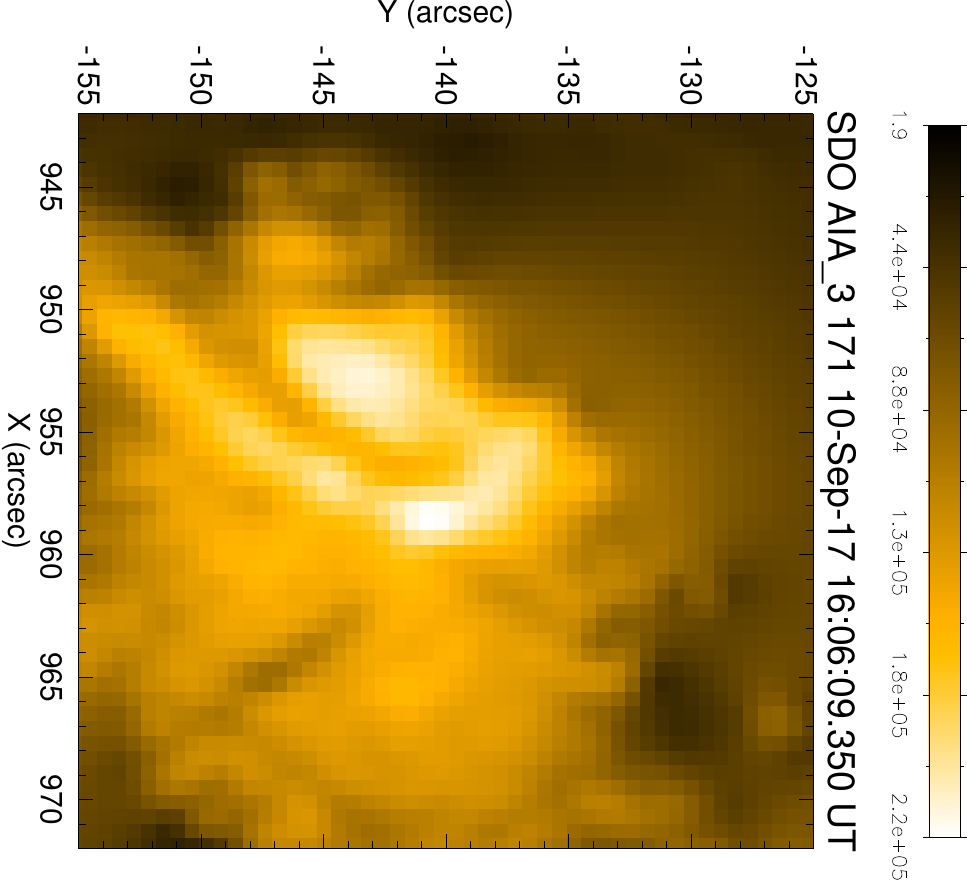}} 
\subfigure[{T16:06:21}]{\includegraphics[width=0.2\textwidth, angle=90]{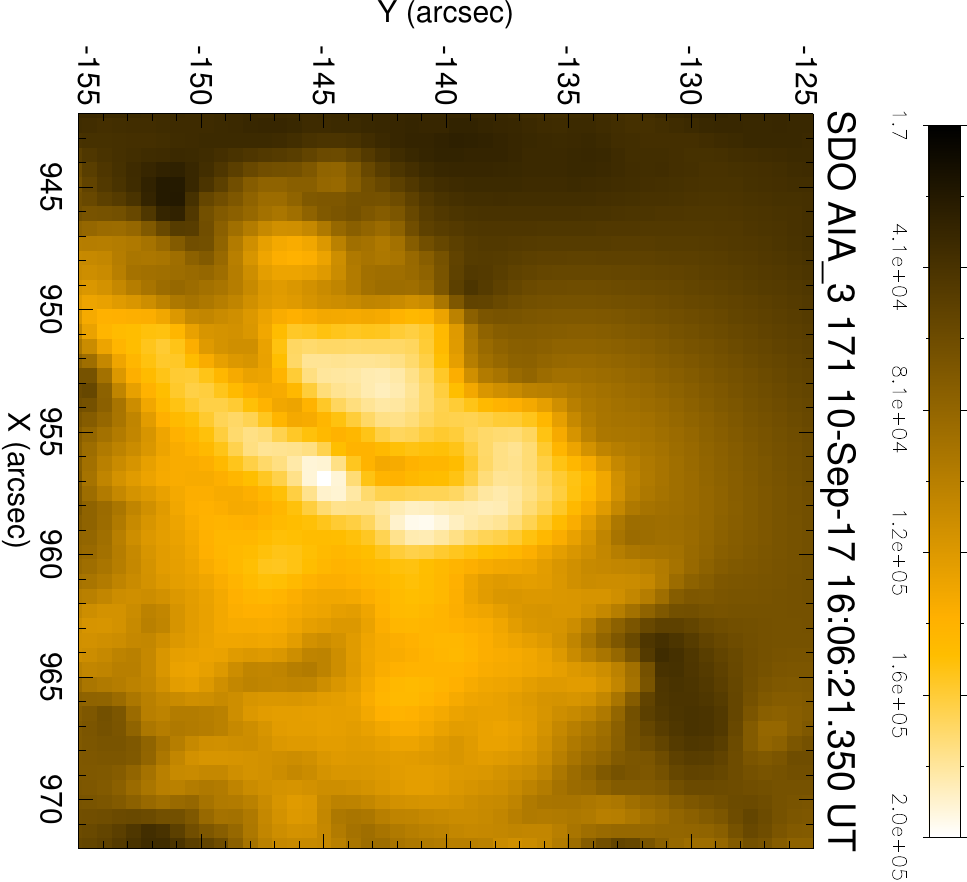}} 
\subfigure[{T16:06:33}]{\includegraphics[width=0.2\textwidth, angle=90]{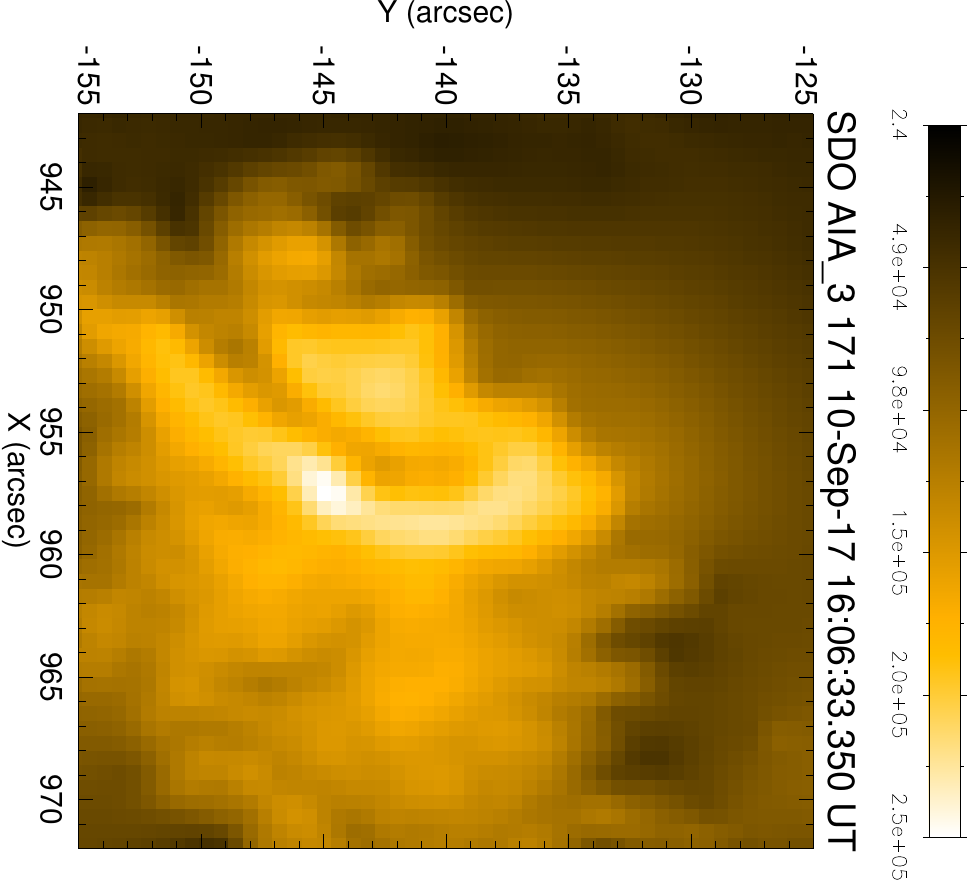}} 
\\
\subfigure[{T16:06:21}]{
\includegraphics[width=0.3\textwidth, angle=90]{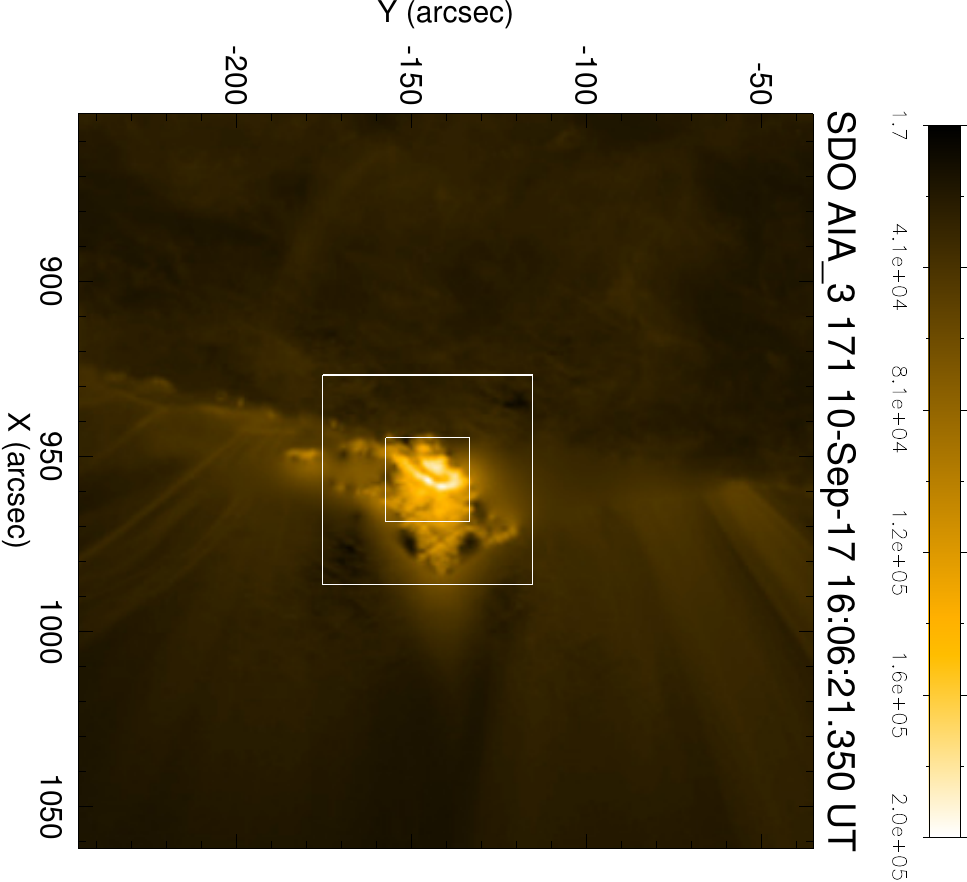}} 
\subfigure[{Counts/duration on time}]{\includegraphics[width=0.4\textwidth]{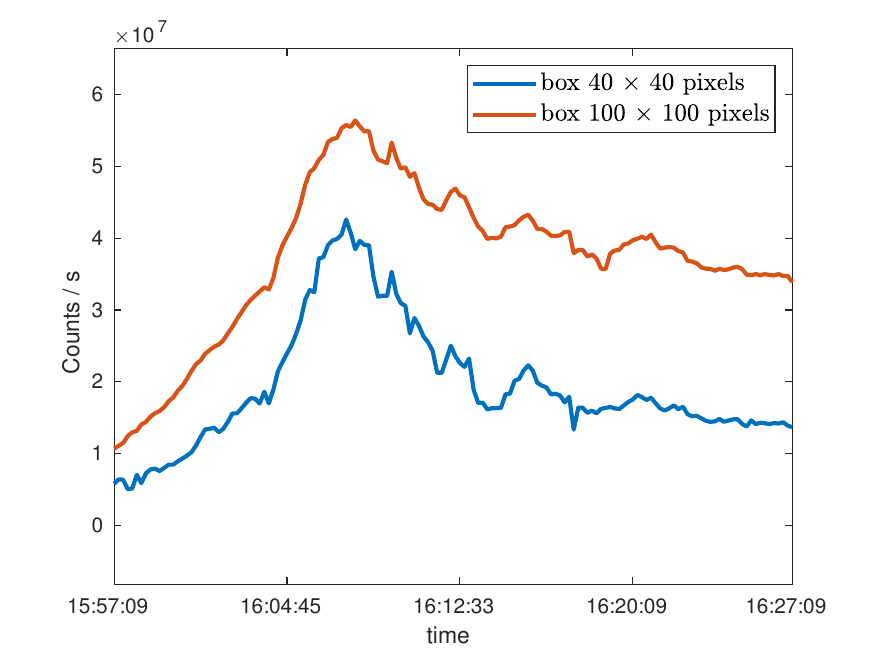}} 
\caption{Bandwidth 171 $\mathring{A}$. First row: saturated images (from 16:05:45 to 16:06:33 UT). Second row: de-saturated images. Third row: zoom of the de-saturated images on the emission core. Fourth row: left panel: de-saturated image at 16:06:21 UT with highlighted the two boxes in which we computed the flux along the acquisition time from 15:57:09 UT to 16:27:09 UT; right panel: reconstructed flux in the two boxes as a function of time.}\label{fig:figure-3}
\end{figure}

\begin{figure}
\subfigure[{T16:00:14 Original}]{
\includegraphics[width=0.30\textwidth, angle=90]{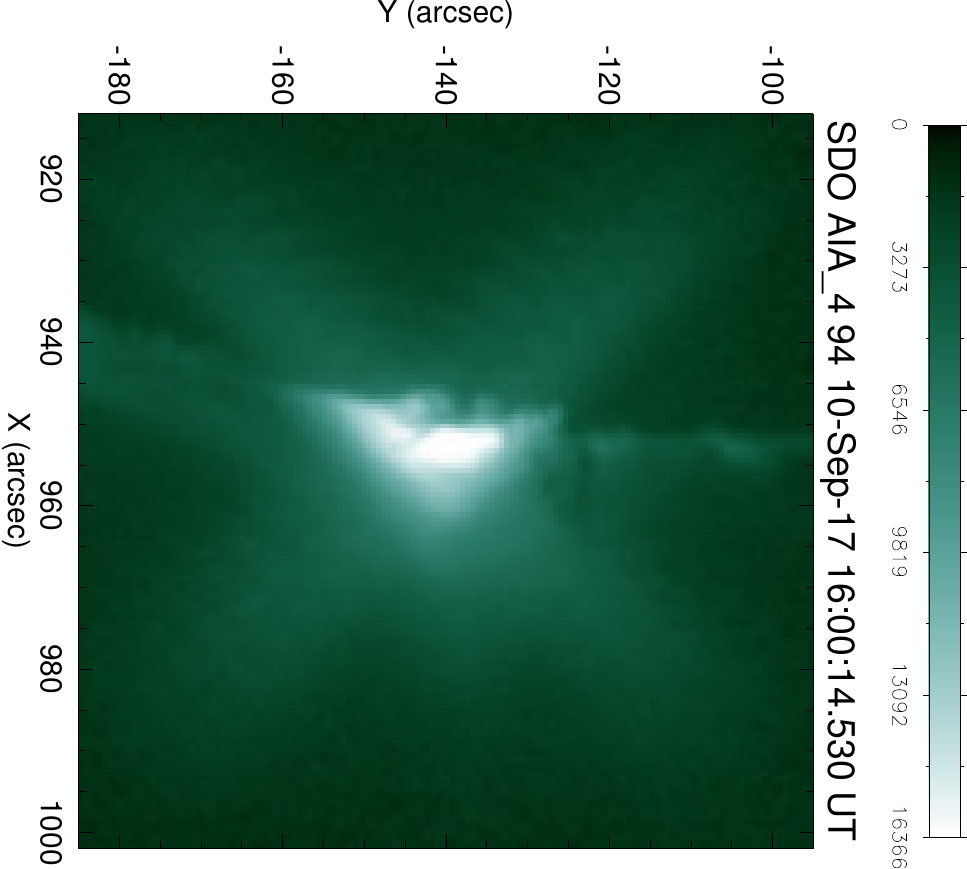}} 
\subfigure[{T16:00:23 Original}]{\includegraphics[width=0.30\textwidth, angle=90]{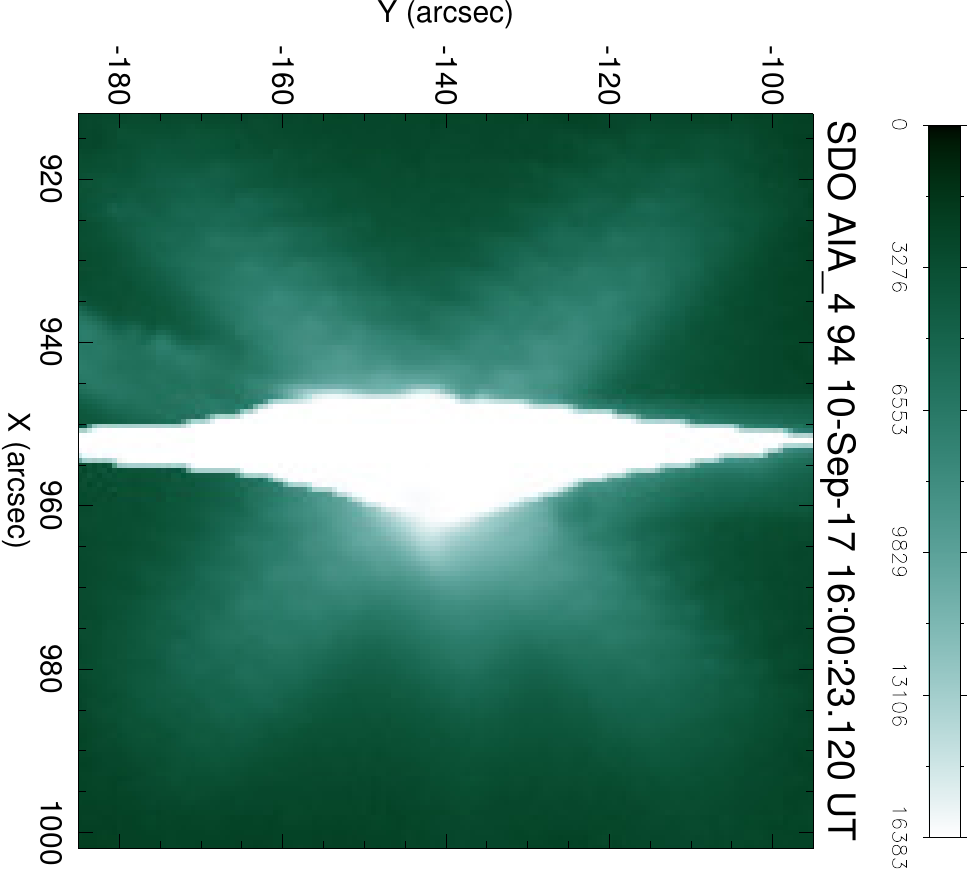}} 
\subfigure[{T16:00:38 original}]{\includegraphics[width=0.30\textwidth, angle=90]{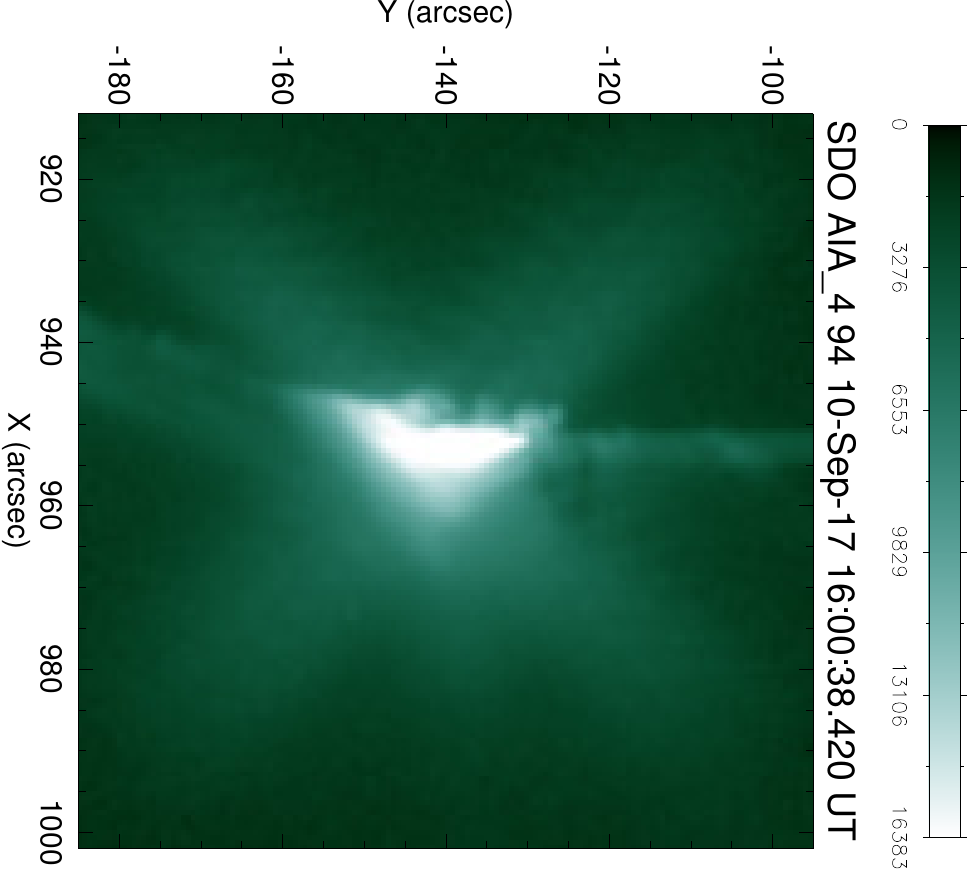}} 
\\
\subfigure[{T16:00:23 Profiles}]{\includegraphics[width=0.3\textwidth, angle=0]{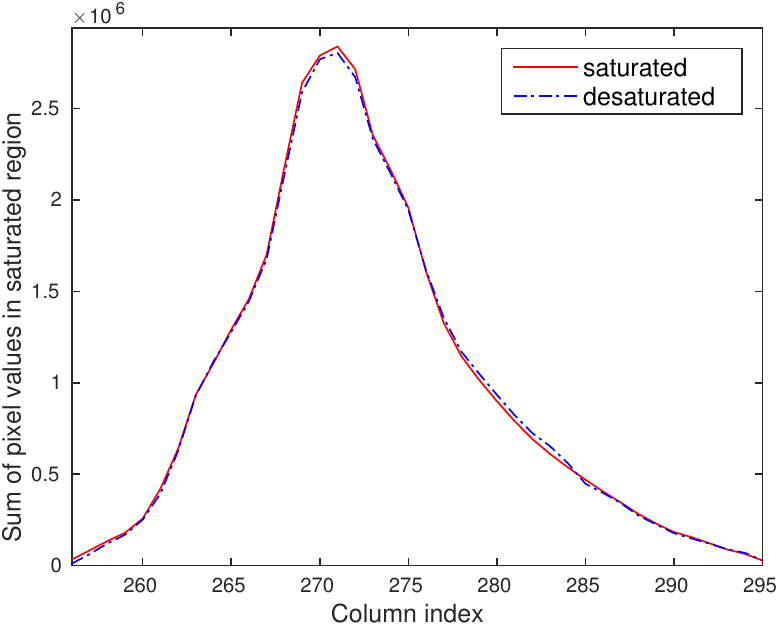}} 
\subfigure[{T16:00:23 Reconstruction}]{\includegraphics[width=0.30\textwidth, angle=90]{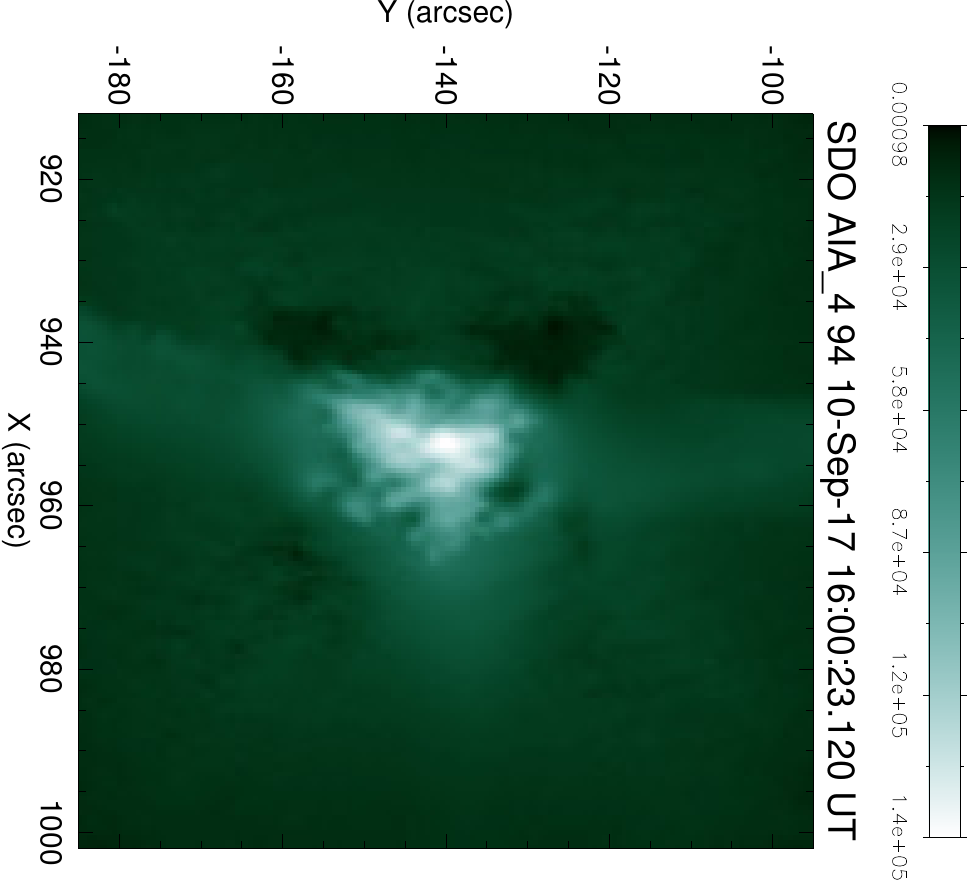}} 
\subfigure[{T16:00:23 Residual}]{\includegraphics[width=0.30\textwidth, angle=90]{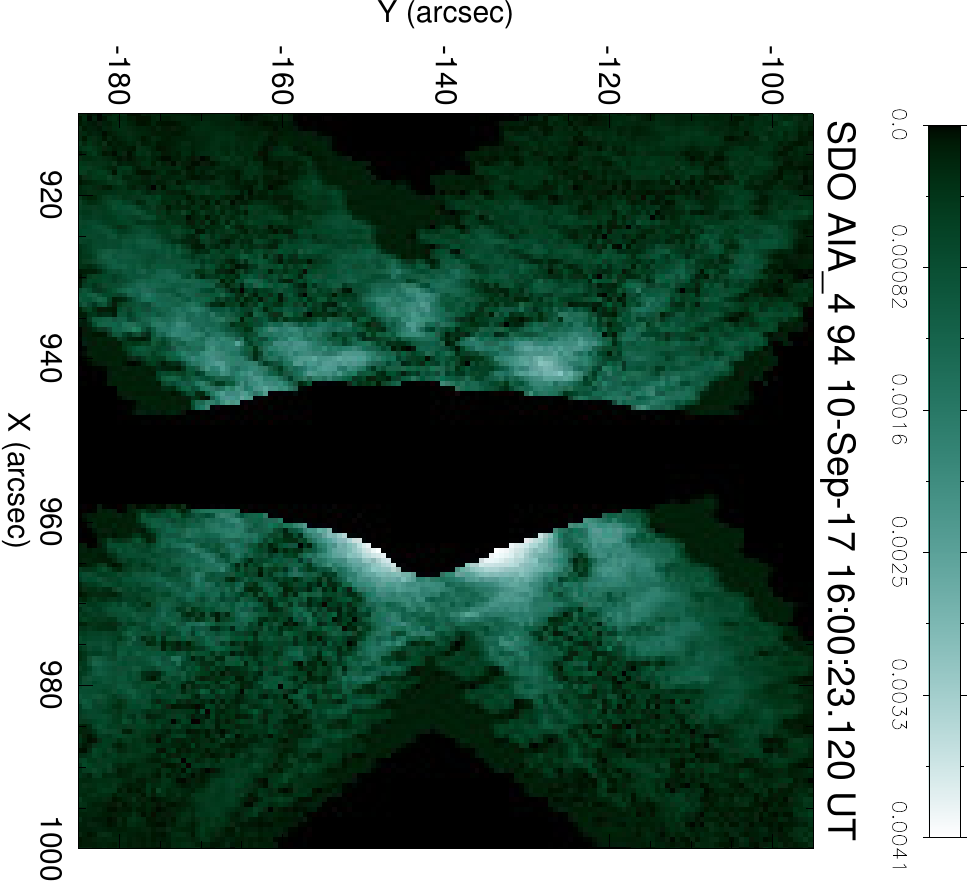}} 
\caption{Bandwidth 94 $\mathring{A}$. First row, middle panel: saturated image at 16:00:23 UT; left and right panels: images captured at 16:00:14 UT and at 16:00:38 UT, respectively. Second row, middle panel: de-saturated image at 16:00:23 UT; left panel: flux in the desaturated image integrated along the image columns corresponding to the saturation region, superimposed to the overall saturated signals in the same columns; right panel: pixel-wise C-statistics in correspondence with the diffraction fringes.}\label{fig:figure-4}
\end{figure}

The desaturation power of our algorithm is illustrated in Figure \ref{fig:figure-3} together with a first example of how these recovered EUV images can be used for basic scientific purposes. The figure refers to the most saturated filter in the batch of {\em{AIA}} wavelengths observing the flaring storm on September 10 2017. Around $300$ images in the time range between 15:45:09 UT and 16:45:09 UT were dramatically corrupted by a wide saturation stripes so that more than one hour observation of one of the most intriguing explosive events in the last twenty-five years could not be utilized for scientific investigation. The first row of the figure shows five consecutive ones of these images in the time range 16:05:45 UT - 16:06:33 UT; the blooming effects are clearly not distinguishable from the primary saturation region while the diffraction fringes affect around half of the remaining field-of-view. These same fringes were given to the algorithm that produced the restored images represented in the second row and zoomed in the third one, where the core of the flare is visible during its temporal evolution. The peak intensity in these cores is more than $10^5$ DN pixel$^{-1}$, which is well above the saturation level of $16383$ DN pixel$^{-1}$. The bottom row of this figure shows that it is now possible to determine the photon flux at $171~ \AA$ over time and that this flux has a realistic behavior with respect to both observations and simulation models \cite{krve15,od10}. The statistical reliability of the desaturated information seems notably good. The C-statistic values in Table \ref{table:table-1} describe the predictive power of the desaturated signal in the primary saturation region when reproducing the experimental diffraction patterns. These numbers are the C-statistic values averaged over the diffraction pixels and corresponding to the desaturation of $50$ highly saturated images in the $171~ \AA$ band: these values go down $1$ at the third iteration of the second step of the algorithm for most images and for all examples the goodness-of-fit is completely satisfactory after just $4$ iterations of the alternate iterative scheme (all desaturated images presented in this paper correspond to the last iteration with C-statistic bigger than $1$). We also applied the algorithm to the processing of images in the $94 ~\AA$ filter, where saturation and blooming effects are typically less significant. The particular case considered in Figure \ref{fig:figure-4} corresponds to a saturated frame at that wavelength, which is recorded immediately before and immediately after two mildly saturated frames (see the panels in the top row of the figure). Comparing the outcome of the desaturation process (middle panel, bottom row) with such mildly saturated maps shows that the morphology of this reconstructed image, corresponding to the $16:00:23$ UT time frame, nicely accommodates along a realistic time evolution from the $16:00:14$ UT time frame (top row, left panel) to the $16:00:38$ UT time frame (top row, right panel). More quantitatively, the plot in the left panel, bottom row, compares the flux integrated along the image columns in all the saturated region, before and after the desaturation process: this quantity has been constrained in the LASSO step of the alternate iterative scheme and, coherently, the fit is satisfactory. Finally, the bottom right panel represents the residuals in the diffraction fringes for the desaturated map: the small pixel values prove that the recovered signal rather satisfactorily fits the experimental data even locally and even in image regions rather far from the places where the diffraction patterns show up in the most significant fashion.

\section*{Discussion}
{\em{SDO/AIA}} is for probably the most powerful instrument for EUV solar imaging ever conceived, opening new crucial windows on the comprehension of how the solar magnetic fields release the huge amount of energy they store, solar storms are triggered at the Sun's surface and they propagate from the solar atmosphere toward Earth. Data analysis is at the core of {\em{AIA}} investigation: this telescope has an unprecedented spatial resolution at all its $10$ wavelengths and the temporal cadence with which it records this highly informative plethora of images requires a processing capability to extract the maximum amount of information with the maximum degree of automation. A big data approach to the analysis of this unprecedented wealth of information at EUV wavelengths has been so far hindered by the presence of saturation effects that significantly degrade the scientific usability of the {\em{AIA}} maps. This deterioration mainly involves the image core and becomes dramatic in the case of GOES X class flare, which are the most interesting ones for the comprehension of the acceleration mechanisms at the basis of energy release in flares. What we showed in the present study is that the information deleted by primary saturation and blooming in the image core is not irremediably lost, but it is all still there, encoded in the peripheral diffraction fringes, that inverse diffraction is able to restore them and accommodate them back into the core, and that this can be done at a single image level, without any need of interpolating pixel contents coming from other unsaturated {\em{AIA}} maps.  What to do next is now clear: thanks to this crucial desaturation step, all ingredients are now at disposal to design and implement an automatic pipeline for a big data processing of {\em{AIA}} production, able to realize the whole stream of operations that from each recorded image leads to a reconstructed EUV map relieved by saturation, diffraction and dispersion effects and therefore ready to a full exploitation within the framework of all possible physical models concerning flaring emission.

\section*{Methods}
{\bf{The imaging model.}} As for all typical telescopes based on a focusing optics technology, {\em{AIA}} images are the result of the convolution between the incoming photon flux and the instrument point spread function (PSF). In the case of {\em{AIA}} the PSF is made of the diffraction component $A_D$ and the diffusion core $A_C$ and therefore {\em{AIA}} records and image $I$ according to the signal formation model
\begin{equation}\label{signal-formation-model}
I = (A_C + A_D)f~,
\end{equation}
where $f$ is the actual incoming photon flux. In particular, if $x^*$ is the signal in the saturated region $S$ (where saturated region here means the sum of the primary saturation pixels and the blooming pixels), the imaging equation at the basis of our desaturation approach reads as
\begin{equation}\label{imaging-equation}
I_F = A_D^S x^* + b~~,
\end{equation}
where $b$ is the signal background in correspondence with the diffraction fringes. Please note that in (\ref{imaging-equation}) $I_F$ is not limited to the diffraction fringes but comprises all image pixels where the signal in $S$ is projected by $A_D$; the use of a sparsity constrain in the desaturation process will automatically allow the segmentation of the true diffraction pattern from the remaining part of the image.

\begin{figure}
\includegraphics[width=0.8\textwidth]{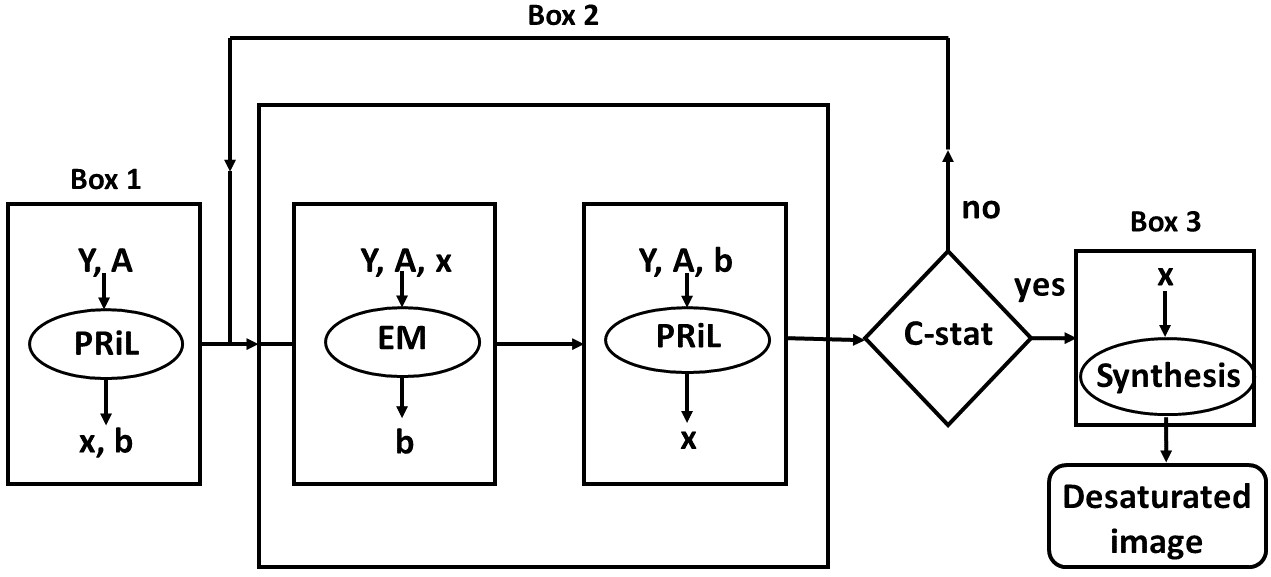} 
\caption{The algorithm scheme.}\label{fig:figure-2} 
\end{figure}

{\bf{The desaturation method.}} In order to illustrate all details of our computational approach, we rely on the scheme described in Figure \ref{fig:figure-2}.
\begin{itemize}
\item Box 1. We compute the constrained PRiL solution of the model equation (\ref{imaging-equation}), i.e. we solve the minimum problem
\begin{equation}\label{min PRiL}  \left\{ \begin{array}{l}
    (\hat{b}, \hat{x}) = \arg\min_{(b,x)\in\mathbb{R}\times \mathbb{R}^{\# S}} \| I_F- A_D^S x - b \|_{PRiL}^2 + \lambda \Vert x \Vert_1 \\
   y=Cx \end{array} \right.
\end{equation}
where $b$ at this step is a constant offset to estimate and $\| \cdot  \|_{PRiL}$ denotes the PRiL approximation of the distance between the observed and predicted data as measured by the discrepancy based on the Poisson statistic. In the constraint $y=Cx$, the $j-$th component of the vector $y$ is the product of the saturation threshold value (i.e., $16383$ DN pixel$^{-1}$) times the number of saturated pixels in the $j$-th column of the image; the matrix $C$ is such that $(Cx)_j = \sum_{i} (A^S_C)_{ji} x_i$, where $i$ runs over the row indexes of the saturated pixels in the $j$-th column. Therefore, this constraint can be easily interpreted as an energy requirement: we impose that the overall energy of the signal in the saturated columns is maintained during the desaturation process. Coherently to this, the regularization parameter $\lambda$ is chosen in such a way that the overall signal provided by the desaturated pixels is equal to the overall signal in the saturated region.
\item Box 2. This illustrates an iterative scheme as follows:
\begin{itemize}
\item input: $I_F$, $A_D^S$, $\hat{x}^{(0)}:= \hat{x}$, $b^{(0)}:=\hat{b}$.
\item $k$-th iteration $k\ge 0$:
\begin{enumerate}[(a)]
\item we estimate the background by computing an iteration of the EM algorithm
\begin{equation}\label{EM}
        b^{(k+1)} = b^{(k)} \frac{I_F}{A_D^S\hat{x}^{(k)}+b^{(k)}}~,
    \end{equation}
where divisions and products must be intended as component-wise.
    \item Then we solve Equation (\ref{imaging-equation}) by using the background estimate $b^{(k+1)}$ provided by the EM step. In details:
\begin{equation}\label{min PRiL}  \left\{ \begin{array}{l}
    \hat{x}^{(k+1)} = \arg\min_{x \in \mathbb{R}^{\# S}} \| I_F- A_D^S x - b^{(k+1)} \|_{PRiL}^2 + \lambda \Vert x \Vert_1 \\
   y=Cx \end{array} \right.
\end{equation}
where $\lambda$ is selected with the procedure described in Box 1 item.
    \item Update background iteration
    \begin{equation}
        b^{(k)} = b^{(k+1)}~.
    \end{equation}
    \item Update solution iteration
    \begin{equation}
        \hat{x}^{(k)}=\hat{x}^{(k+1)}~.
    \end{equation}

    \end{enumerate}
    \item We stop the iterative procedure when the C-statistics computed on fringes is approximately equal to 1. \\
\item  Output: $\hat{x}^{(k_{opt})}$~.
\end{itemize}
Note that EM in Equation (\ref{EM}) is stopped according to the optimal rule formally described in \cite{bepi14} and applied in \cite{beetal13} for the reconstruction of hard X-ray images.
\item Box 3. The outcome $\hat{x}^{(k_{opt})}$ of Box 2 is the desaturated signal in the overall saturation region $S$. In order to segment in it the blooming region $B$ and the primary saturation region $P$, we exploit the fact that the optimization process utilizes a sparsity constraint. Therefore, $B$ is given by all pixels with grey level zero while $P$ is given by all pixels with grey level bigger than zero. This allows a final synthesis step which is given by
\begin{equation}
   I_{\rm desat} = 
   \begin{cases} (A^{S}_C \hat{x}^{(k_{opt})})_{P}, & \mbox{ in } P \\
                    \text{smoothing procedure},  & \mbox{ in } B \\
   I_{F}-A^S_D \hat{x}^{(k_{opt})}, & \mbox{ in } F  \\ I , & \mbox{ elsewhere. } 
   \end{cases}
\end{equation}
In the blooming region we use the automated smoothing procedure introduced in \cite{ga10,waetal12}.
\end{itemize}

\noindent {\bf{Acknowledgements.}}
One of the authors (SG) acknowledges the financial support of a grant from the Italian Gruppo Nazionale di Calcolo Scientifico (GNCS).\\
{\bf{Author contributions.}} SG, FB, and MP formulated the desaturation algorithm. SG wrote the code implementing the algorithm; MP, RS, and AMM contributed to the discussion and interpretation of results. MP wrote the text with the support of SG.

\bibliographystyle{model1a-num-names}
\bibliography{<your-bib-database>}

\begin{table}
\small
\caption{Performance of our method in terms of C-statistics in the case of the September, 10 2017 solar storm,
recorded at waveband 171 $\mathring{A}$
for different acquisition times. 
First and sixth columns: recording time. 
From the second to the fifth columns and from the seventh to the tenth columns: 
C-statistics values at the first four iterations of the algorithm for the corresponding recording time.}
\centering
\begin{adjustbox}{max width=\textwidth}
\begin{tabular}{ l l l l l | l  l l l l l l}
 time (UT)   &   iter 1 & iter 2 & iter 3 & iter 4  & time (UT)   &   iter 1 & iter 2 & iter 3 & iter 4 \\
\midrule                                                                         
16:04:09 & 2.96 & 0.86 &  &  & 16:09:33 & 4.82 & 1.45 & 1.02 & 0.72 \\
\hline                                                                        
16:04:21 & 2.68 & 1.08 & 0.72 &  & 16:09:45 & 5.19 & 1.08 & 0.97 &  \\
\hline                                                                        
16:04:33 & 3.20 & 1.06 & 0.65 &  & 16:09:57 & 4.52 & 1.25 & 0.71 &  \\
\hline                                                                        
16:04:45 & 2.89 & 1.39 & 0.68 &  & 16:10:09 & 4.24 & 1.18 & 0.97 &  \\
\hline                                                                        
16:04:57 & 3.12 & 1.44 & 0.67 &  & 16:10:21 & 3.98 & 1.21 & 0.84 &  \\
\hline                                                                        
16:05:09 & 3.39 & 1.48 & 0.66 &  & 16:10:33 & 4.38 & 1.19 & 0.76 &  \\
\hline                                                                        
16:05:21 & 3.95 & 1.33 & 0.77 &  & 16:10:45 & 3.69 & 0.97 &  &  \\
\hline                                                                        
16:05:33 & 3.69 & 1.49 & 0.80 &  & 16:10:57 & 3.82 & 0.92 &  &  \\
\hline                                                                        
16:05:45 & 3.59 & 1.73 & 0.93 &  & 16:11:09 & 3.50 & 0.90 &  &  \\
\hline                                                                        
16:05:57 & 3.62 & 1.81 & 0.92 &  & 16:11:21 & 3.88 & 1.00 &  &  \\
\hline                                                                        
16:06:09 & 4.36 & 1.61 & 0.84 &  & 16:11:33 & 3.98 & 0.86 &  &  \\
\hline                                                                        
16:06:21 & 4.72 & 1.78 & 0.90 &  & 16:11:45 & 3.82 & 0.90 &  &  \\
\hline                                                                        
16:06:33 & 4.67 & 1.70 & 0.91 &  & 16:11:57 & 3.98 & 1.08 & 0.52 &  \\
\hline                                                                        
16:06:45 & 4.55 & 1.81 & 0.91 &  & 16:12:09 & 4.22 & 1.01 & 0.57 &  \\
\hline                                                                        
16:06:57 & 4.21 & 1.56 & 0.94 &  & 16:12:21 & 4.70 & 1.09 & 0.58 &  \\
\hline                                                                        
16:07:09 & 4.11 & 1.72 & 0.94 &  & 16:12:33 & 4.56 & 1.13 & 0.65 &  \\
\hline                                                                        
16:07:21 & 4.42 & 1.85 & 1.01 & 0.70 & 16:12:45 & 4.56 & 1.03 & 0.72 &  \\
\hline                                                                        
16:07:33 & 4.65 & 1.67 & 0.92 &  & 16:12:57 & 4.38 & 0.98 &  &  \\
\hline                                                                        
16:07:45 & 4.22 & 1.54 & 1.00 & 0.69 & 16:13:09 & 4.23 & 0.87 &  &  \\
\hline                                                                        
16:07:57 & 4.42 & 1.77 & 1.03 & 0.71 & 16:13:21 & 4.18 & 0.85 &  &  \\
\hline                                                                        
16:08:09 & 4.78 & 1.68 & 1.07 & 0.70 & 16:13:33 & 4.07 & 0.79 &  &  \\
\hline                                                                        
16:08:21 & 4.41 & 1.57 & 1.03 & 0.69 & 16:13:45 & 3.73 & 0.71 &  &  \\
\hline                                                                        
16:08:33 & 4.55 & 1.45 & 0.88 &  & 16:13:57 & 3.68 & 0.73 &  &  \\
\hline                                                                        
16:08:45 & 4.30 & 1.29 & 0.87 &  & 16:14:09 & 4.06 & 0.81 &  &  \\
\hline                                                                        
16:08:57 & 4.53 & 1.21 & 0.84 &  & 16:14:21 & 4.25 & 0.89 &  &  \\
\hline\hline
\end{tabular}
\end{adjustbox}
\label{table:table-1}
\end{table}                    
\bibliography{sample.bib}

\end{document}